\documentclass[sigconf]{acmart}

\usepackage{tikz}
\usepackage{tikz-dependency}

\newcommand*\circled[1]{\tikz[baseline=(char.base)]{
            \node[shape=circle,draw,inner sep=.6pt] (char) {#1};}}

\AtBeginDocument{%
  }

\acmConference[ArXiv Preprint]{}{June 03--05,
  2018}{Woodstock, NY}

\begin{document}

\title{Reducing the Footprint of Multi-Vector Retrieval with Minimal Performance Impact via Token Pooling}

\author{Benjamin Clavié}
\email{bc@answer.ai}
\affiliation{%
  \institution{Answer.AI}
  \country{Japan}
}

\author{Antoine Chaffin}
\email{antoine.chaffin@lighton.ai}
\affiliation{%
  \institution{LightOn}
  \country{France}}

\author{Griffin Adams}
\email{ga@answer.ai}
\affiliation{%
  \institution{Answer.AI}
  \country{USA}
}

\renewcommand{\shortauthors}{Clavié et al.}

\begin{abstract}
  Over the last few years, multi-vector retrieval methods, spearheaded by ColBERT, have become an increasingly popular approach to Neural IR. By storing representations at the token level rather than at the document level, these methods have demonstrated very strong retrieval performance, especially in out-of-domain settings. However, the storage and memory requirements necessary to store the large number of associated vectors remain an important drawback, hindering practical adoption. In this paper, we introduce a simple clustering-based token pooling approach to aggressively reduce the number of vectors that need to be stored. This method can reduce the space \& memory footprint of ColBERT indexes by 50\% with virtually no retrieval performance degradation. This method also allows for further reductions, reducing the vector count by 66\%-to-75\% , with degradation remaining below 5\% on a vast majority of datasets. Importantly, this approach requires no architectural change nor query-time processing, and can be used as a simple drop-in during indexation with any ColBERT-like model.
\end{abstract}




\maketitle

\section{Introduction}

Deep-learning based approaches are becoming increasingly popular in retrieval. Effectively, this means using a neural network to assess the relevance of documents for a given query. Many different approaches to neural IR exist, such as single-vector dense representations~\cite{dense}, learned sparse representations~\cite{splade}, or late-interaction methods, such as ColBERT~\cite{colbert}, which employ multi-vector representations by storing a vector for each token in a document.

Using multiple vectors to represent the sequences allows for a more fine-grained and expressive representation compared to compressing all the semantics into a single vector. As a result, this latter approach has frequently been shown to generalize better to out-of-domain settings than dense representations~\cite{interpolation,msshift}. These properties have led to these methods experiencing a considerable amount of interest in the recent months\footnote{Indeed, ColBERTv2 has gone from 40,000 monthly downloads in late 2023 to being one of the most downloaded models on the HuggingFace model hub, with over 5 million monthly downloads. See \url{https://huggingface.co/colbert-ir/colbertv2.0}}.

However, the enhanced expressiveness of multi-vector representations comes at a hefty storage and memory cost. It is relatively trivial to store millions of dense representations using common indexing methods such as HNSW~\cite{hnswindex}. On the other hand, ColBERT and similar approaches require an order of magnitude more vectors per document. Many techniques have been proposed to alleviate this issue: dimension reduction methods through a learned projection layer~\cite{colbert}, aggressive quantization using Inverted File~\cite{ivf}-Product Quantization~\cite{pq} (IVF-PQ) indexing~\cite{colbertv2} and an optimised indexing and querying mechanism~\cite{plaid}.

These methods, while bringing the storage requirement down to the same order of magnitude as single-vector methods, does not fully close the gap with storage and memory requirements of these approaches. This is especially apparent as dense single-vector representations can also be aggressively quantized~\cite{zhan21}, with relatively little performance loss when used in conjunction with quantization-aware training~\cite{cohere}.

Moreover, the indexing methods required to achieve this level of compression with ColBERT add complex processing layers, constraining the index to a specific format. This renders them considerably less flexible, notably in terms of document addition and deletion (\textit{CRUD}) than commonly used dense indexing methods such as HNSW~\cite{hnswindex}. As a result, ColBERT can appear as a worse option for evolving corpora, limiting its practical use.

Another line of work has focused on attempting to reduce the total number of vectors needed to be stored. However, the existing methods have not reached widespread adoption, either due to requiring a modified pipeline and specific training for limited gains~\cite{colberter}; or as a result of minimal storage reduction in comparison to the negative retrieval performance impact~\cite{colbertpruning}.

Recently, a study has also confirmed the intuition that different tokens have vastly varying levels of importance for multi-vector  retrieval performance~\cite{xtr}. This is in line with a series of studies in the adjacent field of Computer Vision, which have introduced \textit{token merging}~\cite{tokenmergingvision}, which consists in averaging the representation of multiple tokens into a single vector. These studies have shown that it is possible to merge tokens without requiring specific training,  greatly improving the throughput of image processing~\cite{tokenmergingvision} or image generation~\cite{tokenmergingsd}, with very minimal impact on task-specific performance.\\

{\textbf{Contributions}} In this work, \circled{1} we introduce a method we call \textsc{Token Pooling}, which requires no specific training nor pipeline or model modification. This method consists in reducing the total number of vectors retained at indexing time, by average pooling token representations through simple clustering approaches. We explore different clustering methods and show that hierarchical clustering yields the best results. On a variety of commonly used evaluation datasets, this method performs very strongly, allowing for storage cost reductions of up to \textbf{50\%} with no performance degradation on average, and storage cost reductions of \textbf{66\%} with fewer than \textbf{3\%} degradation. \circled{2} We then demonstrate that this approach can be coupled with the usual ColBERT quantization pipeline described above, achieving even greater compression results. \circled{3} Finally, we show that the approach also applies to more than just the ColBERTv2 model and the English language, by showing strong results with Japanese variant of ColBERT on Japanese data~\cite{jacolbert}. 

\section{Token Pooling}

To try and mitigate the issues associated with the high number of vectors to store when using multi-vector retrieval methods, we introduce a simple \textbf{Token Pooling} approach. This is applied at indexing time to reduce the effective number of tokens representing a document. It can be used with any pre-trained ColBERT model, without any further training or modifications.

Our core approach is simple and works as a two-step system: First, we devise a way to group individual vectors together, using one of the three clustering methods detailed below. We then apply \textbf{mean pooling} in order to obtain a single vector which contains an average representation of the cluster. The resulting set of pooled vectors serves as our new multi-vector document representation, and the original vectors are discarded. The main intuition behind this approach is the belief that there is considerable redundancy token vectors within as single a document, and that tokens therefore have varying importance~\cite{xtr}. Within this assumption, pooling the most similar ones is unlikely to considerably modify the overall document representation.

To control the level of compression, we introduce a new variable called the \textsc{pooling factor}. This factor is, effectively, a compression factor. For example, a pooling factor of 2 reduces the total number of vectors stored by a factor of 2, i.e. a 50\% reduction.

\subsection{Pooling Methods}

We explore three pooling methods:

\textbf{Sequential Pooling}. This method acts as our baseline does not require any clustering. Tokens are pooled together based on the order in which they appear in the document, from left to right. In this setting, the pooling factor dictates the number of sequential tokens pooled together. We do not use a sliding window, which means that each token is only ever pooled once. This baseline is inspired by the common intuition that the individual meaning of words is greatly influenced by its direct neighbours~\cite{firth, firthrevisited}.

\textbf{K-Means based pooling}. This method uses k-means clustering~\cite{kmeans} based on the cosine distance between vectors. The pooling factor is used in this setting to define the total number of clusters, which is set at {\textsc{initial token count}/\textsc{Pooling Factor} + 1}.

\textbf{Hierarchical clustering based pooling}. This method uses hierarchical clustering~\cite{hierarch_clustering}, again based on the cosine distance between vectors. We use Ward's method~\cite{wards} to produce our clusters, which intuitively would be well-suited for this task, as it would seek to minimize the distance between the original vector and the pooled outputs. Additionally, it has generally been observed to perform well for text data~\cite{wardstext1, wardstext2}.

Using this method, we effectively iteratively merge the vectors that minimize the ward distance. In this setting, the pooling factor is used to define the maximum number of clusters that can be formed, as a constraint on the clustering algorithm. This effectively means that this method will result in \textbf{at most} {\textsc{initial token count}/\textsc{Pooling Factor} + 1} clusters.

\section{Experimental Setting}

\subsection{Implementation}

\textbf{Models} We conduct the vast majority of our experiments with ColBERTv2~\cite{colbertv2}, trained on English MS-Marco~\cite{msmarco}. To assess that our method is not specific to English nor to ColBERTv2, we also conduct a smaller set of experiments on Japanese using a Japanese version of ColBERT, JaColBERTv2~\cite{jacolbert}. \\

\textbf{Clustering} All clustering methods are implemented using existing libraries and widely used implementations. We use SciPy~\cite{scipy} for hierarchical clustering and a simple, PyTorch-based~\cite{pytorch} implementation for k-means clustering. We evaluate clustering on a wide range of pooling factors: 2, 3, 4, 5, 6 and 8. \\

\textbf{Indexing} All experiments are conducted using the official ColBERT implementation to encode both queries and documents. 2-bit quantization and PLAID~\cite{plaid} indexing are also performed with the original codebase. Experiments with non-quantized vectors are conducted using a basic HNSW indexing implementation, via the \textsc{voyager} library. We provide further details on indexing parameters in Appendix~\ref{app:settings}.

\subsection{Evaluation}

\textbf{Data} We evaluate our method on a varied mix of datasets, in order to capture the impact of the approach in different domains. To do so, we select all small to mid-sized datasets from BEIR~\cite{beir}, the most commonly used retrieval evaluation suite, with the exception of ArguAna\footnote{Due to sensibly diverging results between our ArguAna baseline reproduction and reported results in the literature, as well as the less representative nature of ArguAna, as it is an argument mining dataset, we have not included it in this study}, and from LoTTe~\cite{colbertv2}, another benchmark commonly used for multi-vector approaches. We define as "small and mid-sized" any dataset containing fewer than 500,000 documents. We decide on this cutoff in order to appropriately explore a variety of data sizes while keeping down the cost of experimentation and analysis. In total, this results in 6 datasets from BEIR and 3 from LoTTe. For LoTTe datasets, we use the \textit{search} query subset. We evaluate token pooling with uncompressed vectors on only datasets with fewer than 100,000 documents, and compressed 2-bit vectors on all our chosen datasets. 

For the evaluation of JaColBERTv2, we use two commonly used datasets for Japanese retrieval evaluations, with the same evaluation setting as in the original JaColBERT report~\cite{jacolbert}: JSQuAD~\cite{jsquad} and the Japanese split of MIRACL~\cite{miracl}.


\textbf{Metrics} We report all results in terms of relative performance, where 100 represents the score obtained by the same model with no token pooling. For all datasets, the metrics used to compute this relative performance are based on the ones previously used in the literature. Respectively, NDCG@10 for all BEIR datasets, Success@5 for LoTTe datasets and Recall@5 for JSQuAD and MIRACL (Japanese split). All metrics are computed using the \textsc{ranx}~\cite{ranx} library.

\section{Results}
\label{sec:results}

\subsection{Unquantized Results}

\begin{table}[t]
\begin{tabular}{rccccc}
\multicolumn{1}{l}{}                       & \multicolumn{5}{c}{\textbf{BEIR}}                                                                            \\
\multicolumn{1}{l}{}                       & scifact         & scidocs         & nfc             & \multicolumn{1}{c|}{fiqa}            & \textbf{Avg}    \\ \hline
\multicolumn{1}{l}{\textbf{Hierarc.}}      &                 &                 &                 & \multicolumn{1}{c|}{}                &                 \\
\textit{pool 2}                            & \textbf{100.22} & 97.90           & \textbf{102.02} & \multicolumn{1}{c|}{\textbf{102.33}} & \textbf{100.62} \\
\textit{pool 3}                            & 99.99           & 97.00           & 100.71          & \multicolumn{1}{c|}{98.41}           & 99.03           \\
\textit{pool 4}                            & 96.93           & 95.16           & 101.10          & \multicolumn{1}{c|}{94.91}           & 97.03           \\
\textit{pool 6}                            & 92.44           & 87.48           & 96.58           & \multicolumn{1}{c|}{86.17}           & 90.67           \\ \hline
\multicolumn{1}{l}{\textbf{KMeans}}        &                 &                 &                 & \multicolumn{1}{c|}{}                &                 \\
\textit{pool 2}                            & 95.09           & 99.75           & 97.35           & \multicolumn{1}{c|}{97.31}           & 97.38           \\
\textit{pool 3}                            & 93.13           & 99.65           & 96.16           & \multicolumn{1}{c|}{94.90}           & 95.96           \\
\textit{pool 4}                            & 91.62           & 98.17           & 94.36           & \multicolumn{1}{c|}{90.24}           & 93.60           \\
\textit{pool 6}                            & 89.59           & 94.62           & 91.63           & \multicolumn{1}{c|}{84.82}           & 90.166          \\ \hline
\multicolumn{1}{l}{\textit{\textbf{Seq.}}} &                 &                 &                 & \multicolumn{1}{c|}{}                &                 \\
\textit{pool 2}                            & 95.37           & \textbf{103.01} & 96.46           & \multicolumn{1}{c|}{95.41}           & 97.56           \\
\textit{pool 4}                            & 87.80           & 95.91           & 85.78           & \multicolumn{1}{c|}{82.61}           & 88.03          
\end{tabular}
\caption{Relative performance of various token pooling methods, applied to 16-bit unquantized vectors with HNSW Indexing. A score of 100 corresponds to the model performance without any pooling, and all results are relative to it.}
\label{tab:resultsnoquant}
\end{table}

\begin{table*}[t]
\centering
\begin{tabular}{rcccccccc|cccc|c}
\multicolumn{1}{l}{}                       & \multicolumn{8}{c|}{\textbf{BEIR}}                                                                                                                                                                                                                                         & \multicolumn{4}{c|}{\textbf{LOTTE}}                                                                                                                                 & \textbf{Overall}                                                    \\
\multicolumn{1}{l}{}                       & scifact        & scidocs         & nfc            & fiqa           & \begin{tabular}[c]{@{}c@{}}trec\\ covid\end{tabular} & \multicolumn{1}{c|}{Touché}          & \textbf{Avg}                         & \textbf{\begin{tabular}[c]{@{}c@{}}w/o \\ outliers\end{tabular}} & \begin{tabular}[c]{@{}c@{}}Wri-\\ ting\end{tabular} & Recrea.         & \multicolumn{1}{c|}{\begin{tabular}[c]{@{}c@{}}Life-\\ style\end{tabular}} & \textbf{Avg}   & \textbf{\begin{tabular}[c]{@{}c@{}}Avg w/o\\ outliers\end{tabular}} \\ \hline
\multicolumn{1}{l}{\textbf{Hierarc.}}      &                &                 &                &                &                                                      & \multicolumn{1}{c|}{}                & \multicolumn{1}{c|}{}                &                                                                  &                                                     &                 & \multicolumn{1}{c|}{}                                                      &                &                                                                     \\
\textit{pool 2}                            & \textbf{98.11} & 98.76           & \textbf{98.87} & \textbf{98.16} & 102.95                                               & \multicolumn{1}{c|}{112.50}          & \multicolumn{1}{c|}{101.56}          & \textbf{99.67}                                                   & \textbf{99.65}                                      & \textbf{100.61} & \multicolumn{1}{c|}{98.54}                                                 & \textbf{99.60} & \textbf{99.64}                                                      \\
\textit{pool 3}                            & 94.12          & 98.74           & 94.15          & 89.80          & 101.27                                               & \multicolumn{1}{c|}{123.29}          & \multicolumn{1}{c|}{100.23}          & 97.07                                                            & 97.78                                               & 96.07           & \multicolumn{1}{c|}{97.62}                                                 & 97.16          & 97.11                                                               \\
\textit{pool 4}                            & 93.28          & 98.10           & 93.22          & 84.46          & 101.05                                               & \multicolumn{1}{c|}{132.34}          & \multicolumn{1}{c|}{100.41}          & 96.41                                                            & 96.73                                               & 93.04           & \multicolumn{1}{c|}{96.70}                                                 & 95.49          & 96.02                                                               \\
\textit{pool 6}                            & 90.23          & 94.34           & 91.23          & 74.63          & 101.45                                               & \multicolumn{1}{c|}{136.49}          & \multicolumn{1}{c|}{98.06}           & 94.31                                                            & 92.99                                               & 84.72           & \multicolumn{1}{c|}{91.58}                                                 & 89.76          & 92.36                                                               \\ \hline
\multicolumn{1}{l}{\textbf{KMeans}}        &                &                 &                &                &                                                      & \multicolumn{1}{c|}{}                & \multicolumn{1}{c|}{}                &                                                                  &                                                     &                 & \multicolumn{1}{c|}{}                                                      &                &                                                                     \\
\textit{pool 2}                            & 94.18          & \textbf{100.05} & 93.25          & 89.18          & \textbf{103.11}                                      & \multicolumn{1}{c|}{131.23}          & \multicolumn{1}{c|}{\textbf{101.83}} & 97.65                                                            & 97.31                                               & 96.97           & \multicolumn{1}{c|}{\textbf{98.72}}                                        & 97.67          & 97.66                                                               \\
\textit{pool 3}                            & 93.71          & 98.76           & 89.57          & 83.23          & 101.75                                               & \multicolumn{1}{c|}{136.24}          & \multicolumn{1}{c|}{100.54}          & 95.95                                                            & 96.15                                               & 94.10           & \multicolumn{1}{c|}{96.34}                                                 & 95.53          & 95.77                                                               \\
\textit{pool 4}                            & 94.41          & 97.77           & 87.89          & 80.54          & 99.66                                                & \multicolumn{1}{c|}{138.94}          & \multicolumn{1}{c|}{99.87}           & 94.93                                                            & 94.39                                               & 90.92           & \multicolumn{1}{c|}{93.95}                                                 & 93.09          & 94.14                                                               \\
\textit{pool 6}                            & 87.35          & 90.10           & 85.80          & 72.28          & 96.46                                                & \multicolumn{1}{c|}{\textbf{142.13}} & \multicolumn{1}{c|}{95.69}           & 89.93                                                            & 90.42                                               & 81.09           & \multicolumn{1}{c|}{90.51}                                                 & 87.34          & 88.82                                                               \\ \hline
\multicolumn{1}{l}{\textit{\textbf{Seq.}}} &                &                 &                &                &                                                      & \multicolumn{1}{c|}{}                & \multicolumn{1}{c|}{}                &                                                                  &                                                     &                 & \multicolumn{1}{c|}{}                                                      &                &                                                                     \\
\textit{pool 2}                            & 96.55          & 99.34           & 94.07          & 89.75          & 94.67                                                & \multicolumn{1}{c|}{84.16}           & \multicolumn{1}{c|}{93.09}           & 96.16                                                            & 88.90                                               & 93.04           & \multicolumn{1}{c|}{95.79}                                                 & 92.58          & 94.62                                                               \\
\textit{pool 4}                            & 86.47          & 89.29           & 78.44          & 74.13          & 78.48                                                & \multicolumn{1}{c|}{91.20}           & \multicolumn{1}{c|}{83.01}           & 83.17                                                            & 80.49                                               & 80.79           & \multicolumn{1}{c|}{90.66}                                                 & 83.98          & 83.52                                                              
\end{tabular}
\caption{Relative performance of various token pooling methods, applied to 2-bit quantized vector with PLAID Indexing. A score of 100 corresponds to the model performance without any pooling, and all results are relative to it.}
\label{tab:resultsquant}
\end{table*}

\begin{figure}[h]
  \centering
  \includegraphics[width=\linewidth]{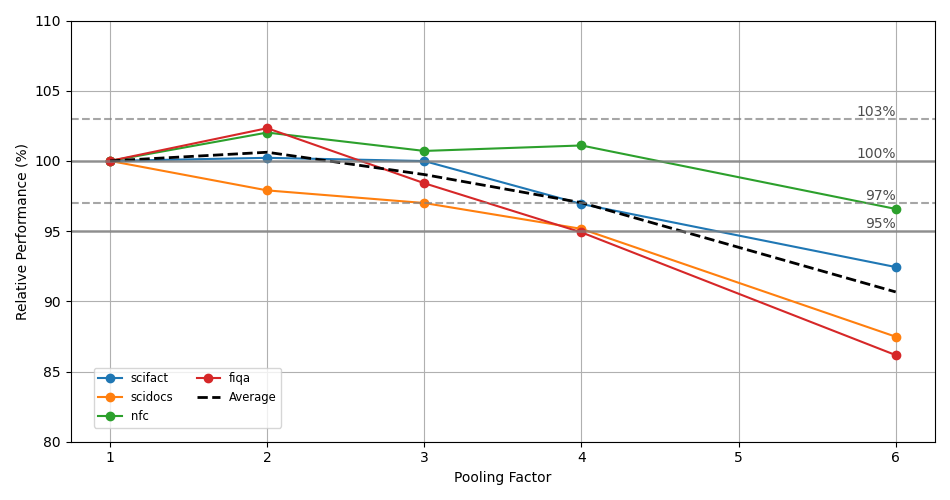}
  \caption{Relative performance degradation at various pooling factors using 16-bit vectors with HNSW indexing.}
  \label{fig:unquant_degradation}
\end{figure}

Table~\ref{tab:resultsnoquant} presents the detailed results for various pooling factors and pooling methods. In the interest of simplicity and space, we only report the results of sequential pooling for factors 2 and 4, as the observed performance degradation varies wildly between datasets and increases is too quickly for this approach to be viable in comparison to the other ones. Worth noting however that, despite its overall noticeably weaker performance, sequential pooling performs remarkably strongly on the \textsc{scidocs} dataset, reaching the strongest performance at a pooling factor of 2, before degrading behind the other two methods at factor 4.
An overview of the relative performance degradation of the best-performing pooling method, hierarchical pooling, across pooling factors on all the small-sized evaluation datasets can be found in Figure~\ref{fig:unquant_degradation} .

We observe that token pooling performs remarkably well in this setting, on all four datasets. In fact, a pooling factor of 2, resulting in a vector count reduction of 50\%, actually slightly \textbf{increases} retrieval performance on average, doing so on 3 out of 4 datasets. A pooling factor of 3 achieves an average performance degradation of less than 1\%, despite reducing storage requirements by over 66\%. We observe a starker degradation from pooling factors of 4 onwards, with an average degradation of 3\% and a degradation of around 5\% on two of the evaluated datasets.

Further compression continues to degrade performance, with a prohibitive average degradation of 10\% at a pooling factor of 6. However, the very strong results of the lower pooling factors show that it is possible to drastically reduce the number of vectors needing to be stored for late-interaction retrieval, at virtually no cost to retrieval performance. These results are especially promising as the lower pool factors offer a comparatively greater gain than subsequent increases.

\subsection{Quantized Results}

The next experiment focuses on evaluating whether this method can be combined with standard ColBERTv2 quantization, enabling its practical use on larger datasets.

\begin{figure}[t]
  \centering
  \includegraphics[width=\linewidth]{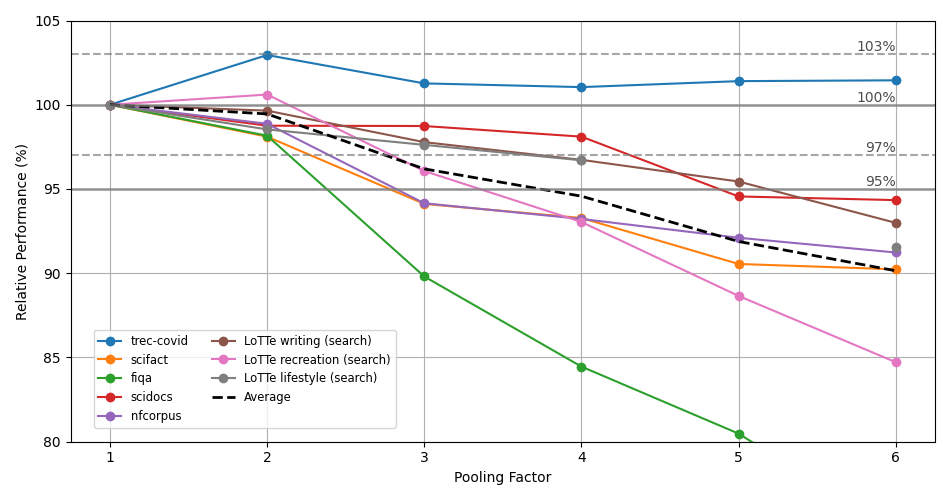}
  \caption{Relative performance degradation at various pooling factors using PLAID-indexed 2-bit vectors. fiqa at factor 6 is truncated for readability.}
  \label{fig:quant_degradation}
\end{figure}

As above, we report the detailed results in Table~\ref{tab:resultsquant}, with the full sequential results truncated, and the overview of relative performance degradation with hierarchical pooling in Figure~\ref{fig:quant_degradation}.

Immediately, we notice one obvious outlier: Touché~\cite{webis}. Not only does performance never decrease with pooling, but it steadily increases, up to a 42.16\%  increase in retrieval performance at a pooling factor of 6. This dataset has frequently been noted as producing unusual results and a recent study has shown that due to its nature as an argument-mining dataset repurposed for retrieval, it is an outlier in many ways and contains considerable noise~\cite{toucheisweird}. As a result, we choose to leave further exploration of this behavior for future work.

There is a second, less-pronounced, outlier in place of fiqa. However, unlike Touché, it does follow a similar trend to other datasets, although the performance degrades noticeably faster as the pool factor increases. While we leave further analysis to future work, fiqa is a highly specialized dataset within the financial domain, and its queries tend to focus on very fine-grained details~\cite{fiqa}, which could partially explain this quicker degradation.

Discounting outliers, we observe overall similar results in the quantized settings as with unquantized vectors. Performance degradation appears to be slightly more pronounced but still very contained, with an average degradation of 1.34\% at a pool factor of 2, and 3.52\% at a pool factor of 3. While nearly all datasets share the same overall trend, there are noticeable variations between datasets, with some beginning to decrease more rapidly at higher pool factors, while others retain performance, even with aggressive pooling.

Interestingly, we note that similarly to the 16-bit setting, KMeans clustering outperforms hierarchical clustering on the \textsc{scidocs} dataset at a pool factor 2, and observe a similar behaviour. However, the performance of hierarchical clustering still noticeably outperforms it on average at every pool factor, and widens the gap at higher pool factors.

\subsection{Vector Count \& Storage Reduction}

Table~\ref{tab:space} provides a comparison in the number of vector stored as well as the disk footprint of storing the full index. We use TREC-Covid at a truncated length of 256 tokens per document (ColBERTv2's default document length~\cite{colbertv2}) for these calculations. We report the number of vectors and index size for 16-bit single-vector dense representations in an HNSW index, as well as PLAID-indexed ColBERT at various pooling factors. Note that the index size reduction is slightly lower than the vector count reduction, due to the indexing overhead introduced by PLAID~\cite{plaid}. 

\begin{table}[h]
\begin{tabular}{r|l}
\multicolumn{1}{l|}{}                           & Index Size \\
\multicolumn{1}{l|}{16-bit Dense Vector}        & 345 MB        \\
\hline
\multicolumn{1}{l|}{\textbf{2-bit PLAID Index}} & 760 MB    \\
\textit{pool 2}                                 & 388 MB    \\
\textit{pool 3}                                 & 260 MB    \\
\textit{pool 4}                                 & 195 MB    \\
\textit{pool 6}                                 & 131 MB   
\end{tabular}
\caption{An overview of index size (in Megabytes) between single-vector representation and PLAID ColBERT indexes.}
\label{tab:space}
\end{table}

We choose to report 16-bit for dense vectors and 2-bit for PLAID, as both methods experience virtually no performance degradation, therefore comparing index without quantization-related performance compromises.

\subsection{Japanese Results}

\begin{table}[h]
\begin{tabular}{rcc|r}
\multicolumn{1}{l}{}                  & JSQuAD               & \multicolumn{1}{l|}{MIRACL} & \multicolumn{1}{l}{Avg} \\ \hline
\multicolumn{1}{l}{\textbf{Hierarc.}} & \multicolumn{1}{l}{} & \multicolumn{1}{l|}{}       & \multicolumn{1}{l}{}    \\
\textit{pool 2}                       & 99.68                & 100.03                      & 99.855                  \\
\textit{pool 3}                       & 98.62                & 97.75                       & 98.185                  \\
\textit{pool 4}                       & 97.69                & 98.26                       & 97.975                  \\
\textit{pool 6}                       & 96.22                & 93.97                       & 95.095                 
\end{tabular}
\caption{Relative performance of hierarchical clustering-based pooling on two Japanese datasets, using quantized 2-bit vectors and PLAID indexing.}
\label{tab:resultsJA}
\end{table}

Finally, Table~\ref{tab:resultsJA} shows the results of token pooling with hierarchical clustering applied to Japanese corpora, using the JaColBERTv2~\cite{jacolbert} model. This evaluation is performed only in the quantized setting, with 2-bit compression on a PLAID~\cite{plaid} index.

While not as thoroughly explored as the English ones, these results show that the pooling approach is neither unique to the ColBERTv2 model nor the English language, as we observe a similar pattern. Performance degradation is minimal at lower pooling factors of 2 and 3, despite the impressive compression.

\section{Conclusion}

In this paper, we introduced \textsc{token pooling}, a simple approach leveraging existing clustering methods and requiring no training nor model modifications to very effectively reduce the numbers of tokens needed to store for multi-vector retrieval models such as ColBERT. 
Our results show that the number of vectors, can be reduced by 50\% with little-to-no performance degradation on the majority of datasets evaluated, thus drastically reducing the size of ColBERT indexes.
Reducing the vector count by 66\% still exhibits minimal degradation, while further compression results in increasing performance degradation.
Notably, this method can lead to ColBERT being applied to a broader range of uses, as it facilitates the use of addition/deletion (\textit{CRUD})-friendly indexing methods such as HNSW.
Our experiments also show that these results hold true even when combined with ColBERT's 2-bit quantization process, allowing for even greater compression than previously possible. We also reproduce our results using a Japanese ColBERT model, showing that this is not limited to a single model, nor to English documents.
With this paper focusing on introducing this method and demonstrating its empirical performance, we hope that our findings will help support future research in better understanding the role of individual tokens in multi-vector retrieval and develop even stronger compression methods.

\begin{acks}
We would like to thank Omar Khattab for his valuable and enthusiastic encouragements while working on this paper.
\end{acks}

\bibliographystyle{ACM-Reference-Format}
\bibliography{bibliography}

\appendix



\section{Retrieval Setting}
\label{app:settings}

The entire retrieval pipeline uses the same approach as the standard ColBERTv2~\cite{colbertv2} + PLAID~\cite{plaid}, including the maxSim scoring function~\cite{colbert}. For querying the PLAID index, we use the best-performing query hyperparameters reported in a recent reproduction study of PLAID~\cite{plaidrepro}: \textit{nprobe}=8, $t_{cs}$=0.3 and \textit{ndocs}=8192.

For the HNSW index, we construct it using generous construction hyperparameters $M=12$ and $EF_{construction}=200$, designed to optimise retrieval performance. At querying time, we use large k values at candidate generation time to ensure the performance is not impacted by the use of approximate search. These hyperparameter choices effectively mean our results are similar to non-approximate searches.

Otherwise, we use the default ColBERTv2 and JacolBERTv2 parameters, with a respective document length of 256 and 300 tokens. We do not perform any particular optimisation step to maximise absolute retrieval performance, as we are focused on the relative performance of different methods. The exact same settings are used for both the baseline runs, and all token pooling methods compared to them.




\end{document}